\documentclass[pra,aps,floats]{revtex4}

\newcommand{\ket}[1]{|#1\rangle}

\begin{document} 


\title{\LARGE \bf Combinatorial Aspects of Nanoscale Magnetism}

\author{W. Florek\thanks{electronic mail: florek@amu.edu.pl},
G. Kamieniarz\thanks{electronic mail: gjk@amu.edu.pl}}

\affiliation{
A. Mickiewicz University, Institute of Physics\\
  ul. Umultowska 85, 61--614 Pozna\'n, Poland}

\author{A. Caramico D'Auria,
U. Esposito, F. Esposito\thanks{electronic mail: filesp@na.infn.it}}

\affiliation{ Universit\`a di Napoli `Federico II', Dipartimento di Scienze Fisiche\\
Piazzale Tecchio, 80125 Napoli and INFM Unit\`a di Napoli, Italy}

\begin{abstract}
 A finite spin system invariant under a symmetry group $G$ is a very 
illustrative example of the finite group action on a set of
mappings $f\colon X\to Y$. In the case of spin systems $X$ is a set of
spin carriers and $Y$ contains $2s+1$ $z$-components $-s\le m\le s$   
for a given spin number $s$. Orbits and stabilizers are
used as additional indices of the symmetry adapted basis. Their mathematical
nature does not lead to smaller eigenproblems, but they label states in 
a systematic way. Some combinatorial and group-theoretical structures,
like double cosets and transitive representations, appear in a natural way. 
In such a system one can construct general formulas for vectors of symmetry
adapted basis and matrix elements of operators commuting with the action of 
$G$ in the space of states. Considerations presented in this paper should
be followed by detailed discussion of different symmetry groups ({\it e.g.}\ 
cyclic of dihedral ones) and optimal implementation of algorithms. The 
paradigmatic example, {\it i.e.}\ a finite spin system, can be useful in
investigation of magnetic macromolecules.
\end{abstract}

\maketitle

\section{Introduction}

Polynuclear clusters provide magnetic materials with a scale intermediate
between that of isolated dimers or trimers and that of bulk materials \cite{nature}.
These magnetic materials exhibit features on a mesoscopic scale, so they may show
quantum effects coexisting with classical behavior. In addition,
large assemblies of spins are interesting as real objects on which one can test
theoretical models with a finite number of spins. Polynuclear magnetic aggregates
have well defined crystal structures \cite{nature}, allowing quantitative comparison
of theoretical results with experimental ones. Unlike other assemblies of small
magnetic particles a typical sample of a molecular magnetic compound is composed
of nominally identical non-interacting magnets with a unique set of chemically
determined parameters. They are complex organometallic systems, too difficult to
investigate them by {\it ab initio}\/ methods, applicable to simple metal cluster.

The aim of this paper is to present and discuss combinatorial 
structure of finite spin models. Such investigation lead to general
analytical formulas for the irreducible basis and matrix elements of 
any operator commuting with all symmetry operators $P(g)$, $g\in G$,
The most important is a Hamiltonian of the system in question, but
also other operators can be taken into account. 
Since $\mathbf{S}^2$ commutes with
the isotropic Heisenberg Hamiltonian, then the total spin $S$ can be
used as an additional label of states. 
Finite spin models have been used in condensed matter physics for many
years \cite{BF} and have been applied to various models of magnetic
materials: from the one-dimensional ferromagnetic Ising model to the
three-dimensional Heisenberg antiferromagnets. 
Recent developments and discoveries have stimulated 
interests in finite spin models, since they can be applied to meso- and
nanoscopic systems, especially magnetic macromolecules with 
well-determined symmetry groups like 
[NaFe$_6$(OCH$_3$)$_{12}$(C$_{17}$H$_{15}$O$_4$)$_6$]ClO$_4$ (Fe$_6$),
[Fe$_{10}$(OCH$_3$)$_{20}$(C$_2$H$_{2}$O$_2$Cl)$_{10}$] (Fe$_{10}$),
Mn$_{12}$acetate or  
[Mn(hfac)$_2$NITPh]$_6$ (Mn$_6$) \cite{aff,andrea,can,jpcm}.
Considering macromolecules with many magnetic centers and large spin numbers 
(up to $s=5/2$) one deals with a very large number of states even for a 
small number of magnetic ions $N$. Moreover, this number increases rapidly 
with increasing $N$ and $s$. However, 
group-theoretical and combinatorial methods yield decomposition of
a space of states into subspaces with relatively small dimensions.
Therefore, magnetic properties of macromolecules can be investigated 
very deeply and model parameters can be fitted in more accurate way.
This paper, based on the results of some earlier papers 
\cite{oldwsf,floirr}, is one of the steps necessary in 
the study considerations of small (nanoscopic) rings
with relatively large spin number. 

It should be also stressed that the approach described and solutions
suggested can be used in any problem where a Hamiltonian 
symmetry group $G$ (or, at least, its subgroup) permutes basis vectors. 
It means that a permutation representation $P$ of $G$ is determined or, in
other words, a group action of $G$ on a basis $\mathcal{B}$ of quantum
space of states is 
introduced. This immediately gives rise to such structures and 
concepts as transitive representations, orbits, stabilizers,
double cosets {\it etc.} Permutation representations themselves
constitute a special case of induced representation and are applied
in many branches of physics \cite{lul84}. Some problems related to
permutation and transitive representations were studied by Lulek 
{\it et al.} \cite{floirr,cow} and their results, in various
extent, are used here.

\section{Finite spin models and simple combinatorial objects}

A system of $N$ spin $s$ gives rise to $(2s+1)^N$ Ising configurations
\[
  |m_1\, m_2\, \dots \, m_N\rangle\,,\qquad m_j\in\{-s,-s+1,\dots,s\}\,.
\]
They can be considered as mappings 
\[
  \mu\colon \{1,2,\dots,N\}\to \{-s,-s+1,\dots,s\}\colon j\mapsto 
  \mu(j)\equiv m_j\,.
\]
Magnetization of the system, $M=\sum_{j=1}^N \mu(j)=\sum_{j=1}^N m_j$,
is an example of an {\em additive weight}. 
Therefore, all combinatorial methods of enumeration with weights can be 
applied \cite{ker}. Moreover, each mapping (configuration) is also characterized
by a {\em content}, {\it i.e.}\/ by numbers
\[
  [k]:=[k_{-s}, k_{-s-1},\dots, k_s]
\]
being numbers of solutions of equations
\[
  \mu(j)=m\,;\qquad m\in \{-s,-s+1,\dots,s\}\,.
\]
In fact $[k]$ is a non-ordered partition of $N$. 
A number of configurations with the same content is given by a polynomial
coefficient
\[
  n([k])=\frac{N!}{\prod_{m=-s}^s k_m!}
  =\frac{(\sum_{m=-s}^s k_m)!}{\prod_{m=-s}^s k_m!}\,.
\]
Hence, a number of configurations (mappings) with a given weight
(magnetization) is
\begin{equation}\label{nM}
  n(N,M)=\sum_{[k]\mid\, \sum jk_j=M} n([k])\,.
\end{equation}
Let 
\[
 \underbrace{D^s\otimes D^s\otimes \dots\otimes D^s}_{N\;\mathrm{times}}
  = \bigoplus_{S=0\;\mathrm{or}\;1/2}^{Ns} n(S)D^S\,.
\]  
be a decomposition of the $N$th tensor power of the (spin) irrep $D^s$. Then
\[
  n(S=Ns)=n(N,M=Ns)=1\,;\quad n(S)=n(N,M=S)-n(N,M=S+1)\;\text{for}
  \; S<Ns\,.
\] 

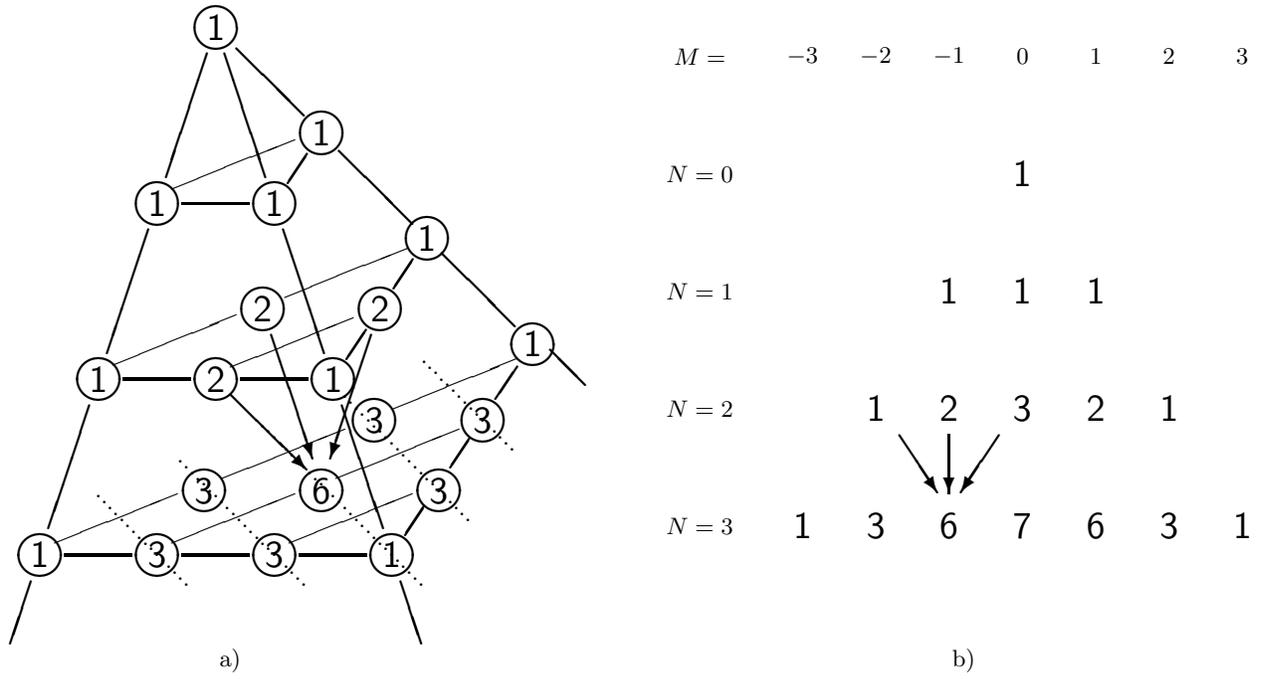
\begin{figure}
\begin{center}
\setlength{\unitlength}{0.39mm}
\begin{picture}(420,235)(0,-15)

\put(75,-6){\makebox(0,0){a)}}
\put(325,-6){\makebox(0,0){b)}}

\thicklines
\put(67,201){\line(-1,-3){14}}
\put(47,141){\line(-1,-3){14}}
\put(27, 81){\line(-1,-3){14}}
\put( 7, 21){\line(-1,-3){ 7}}
\put( 73,201){\line( 1,-3){14}}
\put( 93,141){\line( 1,-3){14}}
\put(113, 81){\line( 1,-3){14}}
\put(133, 21){\line( 1,-3){ 7}}
\put( 77,203){\line( 1,-1){23}}
\put(112,168){\line( 1,-1){25}}
\put(147,133){\line( 1,-1){25}}
\put(184,100){\line( 1,-1){12}}
\put(58.5,150){\line(1,0){23}}
\put(38.5, 90){\line(1,0){23}}
\put(78.5, 90){\line(1,0){23}}
\put(18.5, 30){\line(1,0){23}}
\put(58.5, 30){\line(1,0){23}}
\put(98.5, 30){\line(1,0){23}}
\put( 94.6,156.9){\line(2,3){ 6.6}}
\put(114.6, 96.9){\line(2,3){ 6.6}}
\put(130.6,120.9){\line(2,3){ 6.6}}
\put(134.6, 36.9){\line(2,3){ 6.5}}
\put(150.0, 60.0){\line(2,3){ 6.5}}
\put(165.4, 83.1){\line(2,3){ 7.4}}

\thinlines
\multiput( 60, 20)(-2,2){16}{\makebox(0,0){\small.}}
\multiput(100, 20)(-2,2){22}{\makebox(0,0){\small.}}
\multiput(140, 20)(-2,2){28}{\makebox(0,0){\small.}}
\multiput(156, 42)(-2,2){22}{\makebox(0,0){\small.}}
\multiput(171, 66)(-2,2){16}{\makebox(0,0){\small.}}

\put(55,155){\line(5,2){42}} 
\put(35, 95){\line(5,2){42}}
\put(93.5,118){\line(5,2){42}}
\put( 15, 34){\line(5,2){42}}
\put( 72, 56){\line(5,2){42.2}}
\put(130.5,80){\line(5,2){42.2}}
\put( 75, 94){\line(5,2){42}}
\put( 95, 34){\line(5,2){42}}
\put( 55, 34){\line(5,2){42}}
\put(111, 56){\line(5,2){42}}

\thicklines
\multiput(70,210)(36,-36){4}{\circle{16}}
\multiput(50,150)(-20,-60){3}{\circle{16}}
\multiput(90,150)( 20,-60){3}{\circle{16}}
\put( 86,114){\circle{16}}
\put( 70, 90){\circle{16}}   
\put(126,114){\circle{16}}   
\put( 50, 30){\circle{16}}   
\put( 90, 30){\circle{16}}   
\put(106,52){\circle{16}}
\put(161,76){\circle{16}}
\put(146,52){\circle{16}}   
\put( 66,52){\circle{16}}   
\put(124,76){\circle{16}}   

\put( 70,210){\makebox(0,0){\Large\sf 1}}

\put(106,174){\makebox(0,0){\Large\sf 1}}
\put(142,138){\makebox(0,0){\Large\sf 1}}
\put(178,102){\makebox(0,0){\Large\sf 1}}

\put( 50,150){\makebox(0,0){\Large\sf 1}}
\put( 30, 90){\makebox(0,0){\Large\sf 1}}
\put( 10, 30){\makebox(0,0){\Large\sf 1}}

\put( 90,150){\makebox(0,0){\Large\sf 1}}
\put(110, 90){\makebox(0,0){\Large\sf 1}}
\put(130, 30){\makebox(0,0){\Large\sf 1}}

\put( 86,114){\makebox(0,0){\Large\sf 2}}
\put( 70, 90){\makebox(0,0){\Large\sf 2}}
\put(126,114){\makebox(0,0){\Large\sf 2}}

\put( 50, 30){\makebox(0,0){\Large\sf 3}}
\put( 90, 30){\makebox(0,0){\Large\sf 3}}
\put(161, 76){\makebox(0,0){\Large\sf 3}}
\put(146, 52){\makebox(0,0){\Large\sf 3}}
\put( 66, 52){\makebox(0,0){\Large\sf 3}}
\put(124, 76){\makebox(0,0){\Large\sf 3}}

\put(106, 52){\makebox(0,0){\Large\sf 6}}
\put( 89,105){\vector(1,-3){14}}
\put( 75, 85){\vector(1,-1){26}}
\put(123,105){\vector(-1,-3){14}}
\put(235,200){\makebox(0,0){$M= $}}
\put(235,160){\makebox(0,0){$N=0$}}
\put(235,120){\makebox(0,0){$N=1$}}
\put(235, 80){\makebox(0,0){$N=2$}}
\put(235, 40){\makebox(0,0){$N=3$}}
\put(270,200){\makebox(0,0){$-3$}}
\put(295,200){\makebox(0,0){$-2$}}
\put(320,200){\makebox(0,0){$-1$}}
\put(345,200){\makebox(0,0){$ 0$}}
\put(370,200){\makebox(0,0){$ 1$}}
\put(395,200){\makebox(0,0){$ 2$}}
\put(420,200){\makebox(0,0){$ 3$}}

\put(270, 40){\makebox(0,0){\Large\sf1}}
\put(295, 40){\makebox(0,0){\Large\sf3}}
\put(320, 40){\makebox(0,0){\Large\sf6}}
\put(345, 40){\makebox(0,0){\Large\sf7}}
\put(370, 40){\makebox(0,0){\Large\sf6}}
\put(395, 40){\makebox(0,0){\Large\sf3}}
\put(420, 40){\makebox(0,0){\Large\sf1}}

\put(295, 80){\makebox(0,0){\Large\sf1}}
\put(320, 80){\makebox(0,0){\Large\sf2}}
\put(345, 80){\makebox(0,0){\Large\sf3}}
\put(370, 80){\makebox(0,0){\Large\sf2}}
\put(395, 80){\makebox(0,0){\Large\sf1}}

\put(320,120){\makebox(0,0){\Large\sf1}}
\put(345,120){\makebox(0,0){\Large\sf1}}
\put(370,120){\makebox(0,0){\Large\sf1}}

\put(345,160){\makebox(0,0){\Large\sf1}}

\put(303,71){\vector( 2,-3){13}}
\put(320,71){\vector( 0,-1){20}}
\put(337,71){\vector(-2,-3){13}}
\end{picture}
\end{center}

 \caption{Recurrences for a) polynomial coefficients and b) numbers $n(M)$ for $s=1$
  \label{pas}}
\end{figure}

Some recurrences for $n([k])$ and $n(M)$ can derived from those for 
polynomial coefficients and depicted by analogs of the Pascal triangle. 
In  Fig.~\ref{pas}a the recurrence
  \[
    n([k])=\sum_{j=-s}^s n([k_{-s},k_{-s+1},\dots,k_j-1,\dots,k_s])
  \]
is illustrated by arrows for $n[1,1,1]=n[0,1,1]+n[1,0,1]+n[1,1,0]$. Introducing
this relation to Eq.~(\ref{nM}) one obtains \cite{oldwsf}:
 \[
   n(N,M)=\sum_{m=-s}^s n(N-1,M-m)\,.
 \]
 It is shown by arrows in Fig.~1b in the case $N=3$, $M=-1$: 
$n(3,-1)=n(2,-2)+n(2,-1)+n(2,0)$. The dotted lines in Fig.~1a indicate
polynomial coefficients summed in Eq.~(\ref{nM}). After summation one obtains
numbers presented in the last row in Fig.~1b. Formal equations for the number of
configurations with given $N$ and $M$ can be derived by methods presented
in Kerber's monograph \cite{ker}.

\section{Finite group action}

 For finite both a group $G$ and a set $X$ an action $_GX$ of $G$ on $X$
is defined as a mapping (an external multiplication rule in $X$)
\[
  (G\times X)\to X\colon (g,x)\mapsto gx\qquad \text{with}\qquad
(g_1g_2)x=g_1(g_2)x\qquad \text{and}\qquad 1_Gx=x\,.
\]
A {\em stabilizer} $G_x$ of $x\in X$ is a subgroup of $G$ such that
\[
  G_x:=\{g\in G\mid gx=x\}\,.
\]
An {\em orbit} $G(x)$ of $x\in X$ is a subset of $X$ such that
\[
  G(x):=\{gx \mid g\in G\}\,;\qquad |G(x)||G_x|=|G|\,.
\]

The action $_GX$ can be raised to an action on the set $Y^X$ of 
all mappings $f\colon X\to Y$ ($|Y^X|=|Y|^{|X|}$ for finite sets):
\[
  (G\times Y^X)\to Y^X\colon f\mapsto f'=f\circ g^{-1}\,,\qquad
\mathit{i.e.}\qquad f'(x)=f(g^{-1}x)\,.
\]
Enumeration of mappings needs another subset of $X$ --- the so-called
{\em fixed points}:
\[
  X_g=\{x\in X \mid gx=x \}\,.
\]
 This set is used, for example, to determine number of mappings with given
weight and stabilizer by means of the so-called P\'olya
substitution \cite{ker}.

\section{Transitive representations, double cosets and so on}

Any action of $G$ on an orbit $G(x)$ is similar\footnote{The formal definition of
similarity can be found in \cite{ker}.} to the action
of $G$ on the set 
of left cosets 
\[ 
  G/G_x = \{g_r G_x \mid 1\le r\le |G(x)|\}\,, 
\]
 where $r$ labels representatives $g_r$ of left cosets; this action is 
defined as
 \[
   g(g_rG_x)=(gg_r)G_x\,.
 \]
 If subgroups $G_x$ and $G_y$ are conjugated in $G$, then actions $_G(G/G_x)$
and $_G(G/G_y)$ are similar. It means that nonequivalent actions are labeled
by subgroups $U$ representing classes of conjugated subgroups. These actions,
$_G(G/U)$, can be considered as building blocks of any action $_GX$. 

Let ${\mathcal B}=\{\ket{r}\mid 1\le r\le |G/U|\}$ be a basis of a (formal)
unitary space $L$ constructed as a linear closure (over the field of complex numbers)
of ${\mathcal B}$. The vectors in ${\mathcal B}$ are in a one-to-one correspondence
with elements of an orbit $G(U)$, \textit{i.e.} with left cosets $g_rU$. Moreover,
a representation $P$ of $G$ in the space $L$ can be defined in a natural way \cite{alt}:
 \[
   P(g)\ket{r}=\ket{s}\qquad\text{if}\qquad gg_r\in g_sU\,.
 \]
 This representation is denoted as $R^{G:U}$ and called {\em transitive}. 

A transitive representation $R^{G:U}$ is, in a general case, reducible and
can be decomposed into a direct sum of irreducible representations (irreps) 
of $G$:
\[
 R^{G:U}=\bigoplus n(\Gamma,U)\Gamma\,;\qquad
  0\le n(\Gamma,U)\le [\Gamma]\,,
\]
 where $[\Gamma]$ denotes dimension of $\Gamma$. Any action $_GX$ determines
a permutation representation of $P$ in a space spanned over vectors
$\ket{xr}$, where $x$ labels orbits and $r$ labels representatives of cosets
$g_rG_x$. Since each finite action can be decomposed into action on orbits, then
a permutation representation is decomposed into transitive representations. 

A double coset is determined by two subgroups of $G$ and is defined as
\[
  VgU=\{ vgu \mid v\in V, u\in U\}\,.
\]
Let us consider an action of $G$ on the Cartesian product $G(y)\times G(x)$ 
of two orbits with the stabilizers 
$G_x=U$ and $G(y)=V$. This action is defined as  
 \[ (g,(y,x))\mapsto (gy,gx) \]
 and has the following properties ($x$ and $y$ are fixed):
\begin{itemize}
  \item Each orbit contains an element of the form  $(y,gx)$;
  \item The stabilizer of $(y,gx)$ is $G_y\cap gG_xg^{-1}$;
  \item The mapping
   \( G(y,gx)\mapsto G_ygG_x \)
 is a bijection, so orbits of the introduced action can be labeled
by double coset representatives.
\end{itemize}

\section{Equations for matrix elements}

 It can be shown \cite{cpc,acta} that
any operator ${\mathcal H}$ commuting with all operators $P(g)$, $g\in G$, 
can be written in a quasi-diagonal form with blocks labeled by irreps $\Gamma$.
In each block rows (columns) are indexed by a stability group $U$ ($V$),
representatives $x$ ($y$) of orbits with a stabilizer $U$ ($V$, respectively),
and by indices of vectors $\ket{r}$ ($\ket{s}$) of the irrep $\Gamma$. A general
form of matrix elements is following:
 \begin{equation}\label{eqmat}
   h_{U x r,V y s}(\Gamma)=\left(|U||V|\right)^{-1/2}\:
   \sum_{d=1}^{|V\backslash G/U|} |Vg_dU|
 \langle U\,x\,g_d|{\mathcal H}|V\,y\,e_G\rangle
  B^{UV\Gamma}_{rs}(g_d) \,,
 \end{equation}
 where $B^{UV\Gamma}_{rs}(g_d)$ depends on the so-called reduction coefficients
\cite{lul84} and can be treated as group-theoretical parameters of a model 
under considerations. Analytic formulas for these coefficients can be determined
in the case of cyclic or dihedral group $G$ \cite{cyclic,flonew}.
 This equation simplifies for the unit representation $\Gamma_0$:
 \[
  h_{Ux,Vy}(\Gamma_0)=\left({|U|\over |V|}\right)^{1/2}
   \sum_{j=1}^{|G/U|}
 \langle U\,x\,g_j|{\mathcal H}|V\,y\,e_G\rangle\,,
 \]
 If there is only one double coset, represented by $e_G$, then all left 
coset representatives belong to it and
 \[
  h_{Ux,Vy}(\Gamma_0)=\left(\frac{|G|}{|U||V|}\right)^{1/2}
 \langle U\,x\, e_G|{\mathcal H}|V\,y\,e_G\rangle\,,
 \]
 so it is enough to calculate matrix elements for orbit representatives.
 
 \section{Some remarks on numerical problems}
Application of the formula presented above is not immediate and easy. Of course,
in very simple cases, {\it e.g.}\ $N=4$ and $s=1$, one can calculate all matrix
elements by hand. Larger systems require proper construction and implementation
of algorithms. Moreover, one can use additional quantum numbers (good quantum numbers
as the magnetization and the total spin in the case of isotropic Heisenberg Hamiltonians)
to obtain more detailed classification of states. Secondly, the labeling scheme of
orbit representatives may include partitions of $N$, which, in fact, label 
orbits of the symmetric group $\Sigma_N\supset G$. The action of this group does 
not commute with the Hamiltonian, in a general case, but commutes with the 
operator $S^z=\sum_{j=1}^n s_j^z$ of the total magnetization. Therefore, if one wants
to use the quantum number $M$ as an additional index, then it is natural to include
partitions in the labeling scheme. 

One of advantages of the formula presented is a fact,  that in some cases
(especially for one-dimensional irreps $\Gamma$) obtained matrix elements are
integers. Therefore, considering the eigenproblems for ${\mathbf S}^2$, which
has integer eigenvalues (for $N$ odd and half-integer $s$ one can consider
$4{\mathbf S}^2$ instead), we obtain a set of homogeneous linear equations
with integer coefficients. This set can be solved {\it exactly}, but we have
to cope with very large numbers. This can be done by means of any multiple
precision library, for example {\sf GMP} \cite{gmp}. Prototypes provided by this
library can be also used to store configurations in compact (packed) form.
Moreover, there are as well procedures for high-precision float calculations, so
the Hamiltonian eigenproblem can be solved, too. More detailed discussion
of numerical problems can be found in other papers \cite{cpc,cmst}.

\section{Examples}

The proposed methods are especially useful in the case of
ring-shaped magnetic molecules with cyclic of dihedral symmetry group. 
Below some results for two molecules are presented: 
(i) Ni$_{12}$(O$_2$CMe)$_{12}$(chp)$_{12}$(H$_2$0)$_6$(THF)$_6$ referred to as
Ni$_{12}$ and (ii) Fe$_6$. The first one has the ferromagnetic ground state 
wit the total spin $S=12$ \cite{jpcm}, whereas the second is an example
of a molecular antiferromagnet with $S=0$ in the ground state \cite{aff}.

The Ni$_{12}$ cluster contains a ring of twelve $s=1$ spin carriers. There
are $3^{12}=531441$ Ising configurations. Assuming isotropic interactions
each subspace with a given magnetization $-12\le M\le12$ can be considered 
separately. The largest dimension, 73789, has the subspace with $M=0$. 
Ising configurations spanning this space are labeled by the following non-ordered
partitions of 12 into no more than three parts (number of configurations for 
each partition is given in the parenthesis): [0,12,0] (1), [1,10,1] (132),
[2,8,2] (2970), [3,6,3] (18480), [4,4,4] (34650), [5,2,5] (16632), and 
[6,0,6] (924). In the next step orbit of the symmetric group $\Sigma_{12}$ are
decomposed into orbits of the dihedral group D$_{12}$, what leads to transitive 
representations; they are decomposed into irreps of D$_{12}$. For each irrep 
one can  
\noindent construct a block of the Hamiltonian matrix according with Eq.~(\ref{eqmat}).
The dimensions of these blocks are collected in Table~\ref{tab2}; the largest one 
is 6160 for $\Gamma=E_4$. All eigenvalues (for $M\neq0$, too) of the Hamiltonian
\vspace*{-2pt}
 \[
  {\mathcal H}=-\sum_{j=1}^{12}(J{\mathbf s}_j{\mathbf s}_{j+1}+g\mu_BBs_j^z)
 \]
can be calculated with high numerical
precision and thermodynamic properties can be determined. The best fit to 
experimental results presented by Blake {\it et al.} \cite{blake} is achieved
for the exchange integral $J/k_B=8.5\pm0.5\,K$ and the gyromagnetic ratio 
$g=2.23\pm0.01$ \cite{jpcm}.

The antiferromagnetic ring Fe$_6$ contains six magnetic ions with $s=5/2$, so
there are 46656 Ising configurations. Numbers $n(S)$ and $n(6,M)$ for $M\ge0$
are presented in Table~\ref{nstab}. Considering the ground state one has to take
into account 32 non-ordered partitions. 
Decomposition into orbits of the dihedral group  involves only 
three stabilizers: C$_1$, D$_1$, and D$_3$.
Taking into account decompositions of transitive representations
into irreps one obtains results collected in Table~\ref{ngamma}.

\table
\caption{Multiplicities $n(\Gamma)$ for a ring of $N=12$ spins $s=1$ with 
$G=D_{12}$ and $M=0$}
\label{tab2}
\begin{tabular}{*{10}{c}}
$\Gamma$ & $A_1$ & $A_2$ & $B_1$ & $B_2$ & $E_1$ & $E_2$ & $E_3$ & $E_4$ & $E_5$ \\
$n(\Gamma)$  & 3179 & 2987 & 3107 & 3056 & 6136 & 6158 & 6140 &
 6160 & 6136 
\end{tabular}
\endtable

Our aim is to calculate spin correlation in the ground state assuming 
isotropic Heisenberg interactions. In the considered case ($N=6$, $s=5/2$) the ground
state has symmetry $\Gamma=B_1$, so the dimension of the eigenproblem is 385. However,
we solve it in two steps: (i) at first we determine and orthonormalize eigenvectors
of ${\mathbf S}^2$ operator with the eigenvalue zero; (ii) next we transform the
Hamiltonian matrix to obtain a block labeled by $\Gamma=B_1$ and $M=S=0$. The eigenproblem
for such small matrix ($20\times20$) can be easily solved with high numerical
precision. We 
determine not only eigenvalues, but eigenstates, too, what allows to calculate spin
correlations. The first three low-lying states have energies per spin (in the unit of $J$):
$-7.322$, $-5.761$, $-4.831$, what agrees with the results presented by Lascialfari
{\it et al.} \cite{lgbc}. The spin correlations per bond 
$\omega_z(r)=\langle \sum_j s_j^zs_{j+r}^z\rangle/6$ in these states are following:
\[
\begin{array}{*{5}{c}}
 r=1 && r=2 && r=3 \\  \hline
 -2.440817 && +1.958689 && -1.952410 \\
 -1.920315 && +1.056202 && -1.188439 \\
 -1.610233 &~~& +0.226289 &~~& -0.148779 
\end{array}
\]
 Note that absolute value of the nearest-neighbor $z$-$z$ correlation in the ground state is 
smaller than in the N\'eel state $\ket{N}=\ket{s,-s,s,-s,s,-s}$, when one obtains the value
of 6.25. However, the total spin correlation, equal $3\omega_z(1)$, is stronger
than in the N\'eel state, since $x$-$x$ and $y$-$y$ correlations are equal to zero in this 
state. This fact shows importance of other sates with $M=0$ and follows from quantum
character of the model considered.

\section{Final remarks}

\table
\caption{Numbers $n(S)$ and $n(6,M)$  for $S,M=0,1,\dots,15$
\label{nstab}}

\begin{tabular}{c*{16}{r}}
$S$ or $M$  & 0 & 1 & 2 & 3 & 4 & 5 & 6 & 7 & 8 & 9 & 10 
    & 11 & 12 & 13 & 14 & 15 \\\hline
$n(S)$ & 111 & 315 & 475 & 575 & 609 & 581 & 505 
    & 405 & 300 & 204 & 126 & 70 & 35 & 15 & 5 & 1 \\ 
$(2S+1)n(S)$ & 111 & 945 & 2375 & 4025 & 5481 & 6391 
    & 6565 & 6075 & 5100 & 3876 & 2646 & 1610 & 875 & 405 & 145 & 31 \\ \hline
$n(6,M)$ & 4332 & 4221 & 3906 & 3431 & 2856 & 2247 & 1666
    & 1161 & 756 & 456 & 252 & 126 & 56 & 21 & 6 & 1  
\end{tabular}
\endtable

\table
\caption{Numbers $n(\Gamma)$ for $M=0,1$ and $S=0$; the last one is simply
calculated as the difference of the previous two
\label{ngamma}}

\begin{tabular}{c*{6}{r}}
$\Gamma$ & $A_1$ & $A_2$ & $B_1$ & $B_2$ & $E_1$ & $E_2$ \\ \hline
$n(\Gamma)$, $M=0$ & 385 & 339 & 385 & 339 & 721 & 721 \\
$n(\Gamma)$, $M=1$ & 383 & 325 & 365 & 334 & 699 & 708 \\ \hline
$n(\Gamma)$, $S=0$ &   2 &  14 &  20 &   5 &  22 &  13 
\end{tabular}
\endtable

In this report we present briefly a way from (finite) group action to calculations
of spin correlations. The group-theoretical and combinatorial object appearing
in this course help us to label Ising configurations in the systematic way and
to obtain relatively simple form of the matrix elements. To exploit all possible
simplifications we have assume isotropic interactions. However, there is no limit
imposed on its range: one can easily include next-nearest neighbors or biquadratic
terms $({\mathbf s}_j{\mathbf s}_k)^2$. In the case of uniaxial anisotropy the
total spin $S$ is not a good quantum number, but even in this case thermodynamic 
properties of small magnetic macromolecules can be determined (the total spin
$S$ was not used to obtain results presented in \cite{jpcm}). Such model is frequently
assumed to achieve a better fit to experimental results, for example in the 
case of Mn$_{12}$acetate molecule \cite{can}. The $XYZ$ models, when
the total magnetization is not a good number, are most difficult. However, the 
Ising configurations still can be labeled by partitions and stabilizers, but the
Hamiltonian matrix cannot be decomposed into blocks labeled by $M$; of course the
label $\Gamma$ can be used as long as the Hamiltonian commutes with all operators 
$P(g)$. It should be also stressed that some results presented are obtained using
procedures included in the {\sf GMP} package. For example, we are able to solve
{\it exactly}\/ a set of homogeneous linear equation with integer coefficients.
The methods proposed has been applied to magnetic macromolecules, however 
calculations in the thermodynamic limit are also very important, since
there are still some open problems related with linear magnets
({\it e.g.} critical exponents near $T=0$\,K in the case of ferro-\linebreak
magnetic ordering \cite{lechu}).

\section*{Acknowledgments}

This work is partially supported by the Committee for Scientific Research
under the KBN grant No.\ 2~P03B~074~19. The numerical calculations were
mainly carried out in the the Supercomputing and Networking
Center in Pozna\'n.

\end{document}